\documentclass[a4paper]{jpconf}
\usepackage{graphicx}
\usepackage{amsmath}
\usepackage{slashed}
\usepackage{multirow}
\begin{document}
\title{Vector- and Pseudoscalar-baryon coupled channel systems}

\author{K.~P.~Khemchandani$^{1\dagger}$, A.~Mart\'inez~Torres$^2$, H.~Kaneko$^1$, H.~Nagahiro$^3$, A.~Hosaka$^1$}

\address{$^1$Research Center for Nuclear Physics, Osaka University, Ibaraki 567-0047, Japan.}
\address{$^2$ Yukawa Institute for Theoretical Physics, Kyoto University, Kyoto 606-8502, Japan.}
\address{$^3$Department of Physics, Nara Women's University,  Nara 630-8506, Japan.}

\ead{$^\dagger$kanchan@rcnp.osaka-u.ac.jp}

\begin{abstract}
In this manuscript, I will report the details of our recent work on the vector meson-baryon (VB) interaction,  which we studied with the motivation of finding dynamical generation of resonances in the corresponding systems. We started our study by building a formalism based on the hidden local symmetry and calculating the leading order contributions to the scattering equations by  summing the diagrams with: (a) a vector meson exchange in the $t$-channel (b) an octet baryon exchange in the $s$-, $u$-channels and (c) a contact interaction arising from the part of the vector meson-baryon Lagrangian which is related to the anomalous magnetic moment of the baryons. We find the contribution from all these sources, except the $s$-channel, to be important. The amplitudes obtained by solving the coupled channel Bethe-Salpeter equations for the systems with total strangeness zero, show generation of one isospin 3/2, spin 1/2 resonance and three isospin 1/2 resonances: two with spin 3/2 and one with spin 1/2. We identify these resonances with  $\Delta$ (1900) $S_{31}$, $N^*$(2080) $D_{13}$, $N^*$(1700) $D_{13}$, and $N^*$(2090) $S_{11}$, respectively.  

We have further extended our study by including pseudoscalar meson-baryon (PB)  as the coupled channels of VB systems.  For this, we obtain the PB $\rightarrow$ VB amplitudes by using the Kroll-Ruddermann term where, considering the vector meson dominance phenomena, the photon is replaced by a vector meson. The calculations done within this formalism reveal a very strong coupling of the VB channels to the low-lying resonances like  $\Lambda$(1405) and $\Lambda$(1670), which can have important implications on certain reactions producing them. In addition to this, we find that the effect of coupling the higher mass states to the lighter channels  is not restricted to  increasing the width of those states, it can be far more strong.
\end{abstract}

\section{Introduction}
The vector mesons, which are understood as $(q\bar{q})_{l=1}$ composite systems, have attracted a lot of attention historically, beginning from the observation of the universality of the $\rho$-meson couplings to the light hadrons, which can be understood in terms of the fluctuation of a photon to a vector meson that subsequently interact with the hadrons. Further findings, like the dominance of the the vector mesons in explaining the electromagnetic form factors of the hadrons \cite{sakurai} stimulated the formulation of a theory where the vector mesons  are treated as the dynamical gauge bosons of the hidden local symmetry of the non-linear sigma model \cite{bando1,bando2,bando3}. The resulting theory has been widely used in hadron physics involving the vector mesons (see Refs.~\cite{hls1,hls2,hls3,hls4,hls5,hls6,hls7} as examples).

\begin{figure}[h]
\begin{center}
\includegraphics[width=14cm]{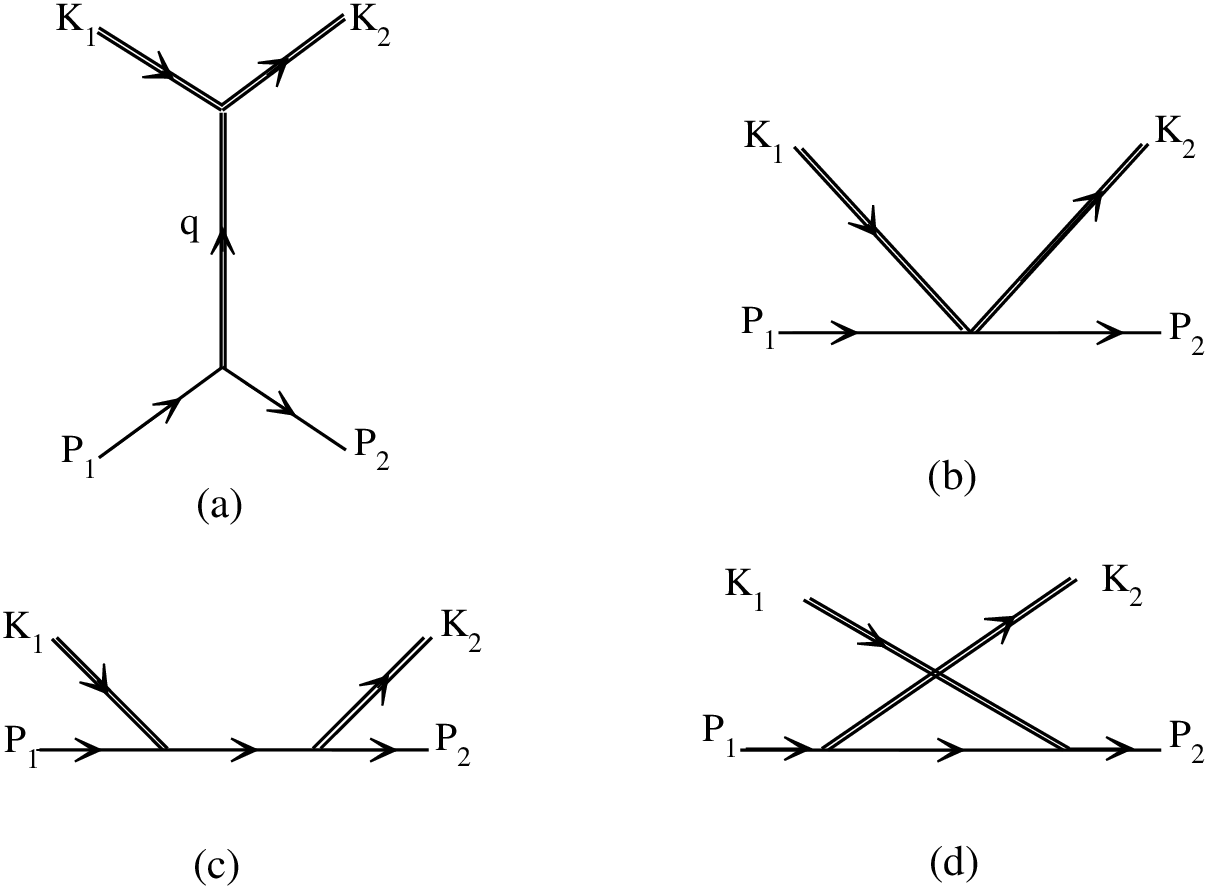}
\end{center}
\caption{\label{fig1} Different diagrams contributing to the vector meson-baryon interactions (see details in the text).}
\end{figure}

Using this effective field theory based on the hidden local symmetry, we have focussed on understanding the low-energy interaction of vector mesons with  baryons, with the idea of finding dynamical generation of baryon resonances, motivated by the large branching ratios of some of the excited baryons to the vector meson-baryon decay channels. Generation of resonances in vector meson-baryon systems has earlier been reported in several phenomenological studies also \cite{juelich,lutz,pedro,michael,sourav,eulogio}.  We proceeded with our study by reminding ourselves that in case of the vector meson-baryon interaction, we cannot rely on the low energy theorems which have worked well in case of the pseudoscalar mesons (see, for example, Refs.~\cite{dgr1,dgr2,dgr3,dgr4,dgr5,dgr6,dgr7} where several baryon resonances have been found to get generated in the pseudoscalar-baryon or even in two/three pseudoscalar-baryon systems). Keeping this  in mind, we built a formalism \cite{us1} on the basis of the hidden local symmetry and calculated the four diagrams shown in Fig.~\ref{fig1}, which correspond to: (a) a vector meson exchange in the $t$-channel, (b) a contact interaction demanded by the gauge invariance (corresponding to the hidden local symmetry (HLS)) of the vector meson-baryon interaction Lagrangian, which can be written for the SU(2) case as:
\begin{equation}
\mathcal{L}_{\rho N } = - g \bar{N} \left\{ F_1 \gamma_\mu \rho^\mu + \frac{F_2}{4M} \sigma_{\mu\nu} \rho^{\mu\nu} \right\} N, \label{rhoNL}
\end{equation}
 and (c, d) the $s$- and $u$-channel octet baryon exchange. 

Writing  the HLS gauge invariant Lagrangian given by Eq.~(\ref{rhoNL}), explicitly, we get  \cite{us1}
\begin{equation}
\mathcal{L} = - g \bar{N} \left( \gamma^\mu  \rho_\mu + \frac{\kappa_\rho}{4M} \left[ \partial_\mu \rho_\nu - \partial_\nu \rho_\mu + i g \left( \rho_\mu \rho_\nu -  \rho_\nu \rho_\mu \right) \right]  \sigma^{\mu\nu}\right) N, \label{one}
\end{equation}
where the nucleon field transforms as $N \rightarrow h(x)N$ with $h(x)$ being an element of the hidden local symmetry, $F_1(q^2 = 0) = 1$, $F_2(q^2 = 0) = \kappa_\rho \sim$ 3.22 is the anomalous magnetic coupling for the $\rho NN$ vertex, $M$ is the mass of the nucleon and 
\begin{equation}
\rho_\mu = \frac{1}{2} \vec{\tau} \cdot \vec{\rho_\mu} = \frac{1}{2} \left(\begin{array}{cc}\rho_\mu^0&\sqrt{2} \rho_\mu^+ \\ \sqrt{2} \rho_\mu^-&-\rho_\mu^0\end{array}\right).
\end{equation}
Using  Eq.~(\ref{one}), we can write  Lagrangians for the vertices involving one and two meson fields, respectively, as
\begin{eqnarray}
\mathcal{L}_{\rho NN}& = & - g \bar{N} \left( \gamma^\mu  \rho_\mu + \frac{\kappa_\rho}{4M} \left( \partial_\mu \rho_\nu - \partial_\nu \rho_\mu  \right) \sigma^{\mu\nu} \right) N \label{3fields}\\
\mathcal{L}_{\rho \rho NN}& = &- ig^2 \bar{N} \left( \rho_\mu \rho_\nu -  \rho_\nu \rho_\mu   \right)\sigma^{\mu\nu} N,
\end{eqnarray}
where the former is used to obtain the amplitudes for the $s$- and  $u$-channel diagrams  \cite{us1}
\begin{eqnarray}
V^{s}_{\rho N} &=& \frac{g^2}{4} \left[ 1 - \dfrac{\kappa_\rho m}{2M}\right]^2 \dfrac{1}{m + 2M}\left(1 - 2 \vec{t_\rho} \cdot \vec{t_N} \right) \left(1 - 2 \vec{s_\rho} \cdot \vec{s_N} \right), \label{vrhon_s}\\
V^{u}_{\rho N} &=&  \frac{g^2}{4}  \left[ 1 + \dfrac{\kappa_\rho m}{2M}\right]^2 \dfrac{1}{m  - 2M}  \left(1 + 2 \vec{t_\rho} \cdot \vec{t_N} \right) \left(1 + 2 \vec{s_\rho} \cdot \vec{s_N} \right), \label{vrhon_u}
\end{eqnarray}
and latter one gives a contact interaction with the amplitude given by
\begin{equation}
V^{CT}_{\rho N} =   \dfrac{g^2 \kappa_\rho}{M} \vec{t_\rho} \cdot \vec{t_N} \vec{s_\rho} \cdot \vec{s_N}.\label{vrhon_ct}
\end{equation}
It should be emphasized here that this contact interaction arises from the $\left[ \rho_\mu, \rho_\nu\right]$ term of the vector tensor, which is essentially required to obtain the HLS gauge invariance of the Lagrangian given by Eq.~(\ref{one}).  

To calculate the $t$-channel diagrams, together
with Eq. (\ref{3fields}), 
we also need
\begin{equation}
\mathcal{L}_{3\rho} = - 2 i g Tr \left(\partial_\mu V_\nu V^\mu V^\nu - \partial_\mu V_\nu V^\nu V^\mu \right),
\end{equation}
which comes from the non-Abelian kinetic term of the $\rho$-meson. The resulting $t$-channel amplitude has the form \cite{us1}
\begin{eqnarray}
V^t_{\rho N} = &-\dfrac{1}{2 f_\pi^2} \left( K_1^0 + K_2^0 \right) \vec{\epsilon_1} \cdot \vec{\epsilon_2} & \,\,\,\, {\rm for  \,\,{\it I} = 1/2} \label{vrhon_t1}\\
= &\dfrac{1}{4 f_\pi^2} \left( K_1^0 + K_2^0 \right) \vec{\epsilon_1} \cdot \vec{\epsilon_2} & \,\,\,\, {\rm  for \,\,  {\it I} =  3/2}, \label{vrhon_t2}
\end{eqnarray}
which  is  spin independent in nature. However  the contact, s- and u-channel
interactions lead to spin dependent amplitudes. A comparison of Eqs.~({\ref{vrhon_s}), ({\ref{vrhon_u}), ({\ref{vrhon_ct}), (\ref{vrhon_t1}) and (\ref{vrhon_t2})  shows that the contribution from all the four diagrams is equally important  (it can be easily seen by using the KSRF relation: $g = m_\rho/\sqrt{2} f_\pi$).  We have calculated the amplitudes for other VB channels using a more general Lagrangian  \cite{us1}, which we will give in the next section. In the subsequent section, we will review the results of the full coupled channel scattering equations for the total strangeness zero VB systems. 

Having built a reliable formalism for the vector-meson baryon interactions, we extended our study by  including pseudoscalar-baryon systems as coupled channels \cite{us2}. Since several resonances decay with large branching ratios to both PB and VB channels, it is possible, then, that a coupled PB-VB dynamics can play important role in explanation of the properties of some of such resonances. However, since  most PB and VB channels lie  well separated on the energy scale, we did not know if we were going to get important information by coupling them. Hence, we  decided to concentrate on studying the low-lying resonances alone since those are better understood. This also allowed us to rely on the $t$-channel calculations of PB $\rightarrow$  PB and VB $\rightarrow$ VB vertices, which simplified the formalism. However, as we will discuss in Section~3, it turned out that the coupling between  the PB and VB channels can result in very useful and significant findings, implying the requirement of a more detailed calculations.

\section{Formalism}

\subsection{Vector-baryon interaction}
To calculate the VB transition amplitudes, we generalized  the SU(2) vector meson-baryon Lagrangian of Eq.~(\ref{one}) to SU(3) to get
\begin{eqnarray} \nonumber
\mathcal{L}_{VBB}&=& -g \biggl\{ \langle \bar{B} \gamma_\mu \left[ V_8^\mu, B \right] \rangle + \frac{1}{4 M} \Bigl( F \langle \bar{B} \sigma_{\mu\nu} \left[ \partial^{\mu} V_8^\nu - \partial^{\nu} V_8^\mu, B \right] \rangle \Bigr.
\biggr.   \\ \nonumber
&+&\Bigl.  D \langle \bar{B} \sigma_{\mu\nu} \left\{ \partial^{\mu} V_8^\nu - \partial^{\nu} V_8^\mu, B \right\} \rangle\Bigr)
+ \langle \bar{B} \gamma_\mu B \rangle  \langle  V_0^\mu \rangle  \\
&+&\left.  \frac{ C_0}{4 M}  \langle \bar{B} \sigma_{\mu\nu}  V_0^{\mu\nu} B  \rangle\right\},\label{yukawaL}
\end{eqnarray}
where the constants $D$ = 2.4 and  $F$ = 0.82. These values were found to well reproduce the magnetic moments of the baryons in Ref.~\cite{jido}.
Further, in our normalization scheme,
\begin{eqnarray}
V =\frac{1}{2}
\left( \begin{array}{ccc}
\rho^0 + \omega & \sqrt{2}\rho^+ & \sqrt{2}K^{*^+}\\
&& \\
\sqrt{2}\rho^-& -\rho^0 + \omega & \sqrt{2}K^{*^0}\\
&&\\
\sqrt{2}K^{*^-} &\sqrt{2}\bar{K}^{*^0} & \sqrt{2} \phi 
\end{array}\right)
\end{eqnarray}
and
\begin{eqnarray}
B =
\left( \begin{array}{ccc}
 \frac{1}{\sqrt{6}} \Lambda + \frac{1}{\sqrt{2}} \Sigma^0& \Sigma^+ & p\\
&& \\
\Sigma^-&\frac{1}{\sqrt{6}} \Lambda- \frac{1}{\sqrt{2}} \Sigma^0 &n\\
&&\\
\Xi^- &\Xi^0 & -\sqrt{\frac{2}{3}} \Lambda 
\end{array}\right),
\end{eqnarray}
which give the octet ($V_8$), and the singlet ($V_0$) fields of the vector mesons by considering the $\omega-\phi$ mixing into account. For this, we write the 
wave functions of the $\omega$- and the $\phi$-meson, under the ideal mixing assumption, as
\begin{eqnarray}\nonumber
\omega &=& \sqrt{\dfrac{1}{3}} \, \omega_8 + \sqrt{\dfrac{2}{3}} \,\omega_0 \\
\phi &=&  -\sqrt{\dfrac{2}{3}}  \,\phi_8 + \sqrt{\dfrac{1}{3}} \,\phi_0.
\end{eqnarray}

With these inputs, we obtained the amplitudes for different processes involving all the VB channels. All these amplitudes are listed in all details in Ref.~\cite{us1}.

\subsection{Coupling pseudoscalar- and vector meson-baryon systems}
The new information required to carry out a coupled channel study of the PB-VB systems is to write the PB $\leftrightarrow$ VB vertex consistently.
This is done by using the Kroll-Ruderman (KR) theorem to write the Lagrangian for the $\gamma N \rightarrow \pi N$ process and by replacing the $\gamma$ by a vector meson via the  notion of the vector meson dominance. More concretely, we write the $\pi N$ Lagrangian from the Gell-Mann-Levi's linear sigma model,
\begin{equation}
\mathcal{L}_{\pi N} = \bar{\psi} \left[ i \gamma^\mu \partial_\mu - g_{\pi NN} \left( \sigma + i \vec{\tau}.\vec{\pi} \gamma_5 \right)  \right] \psi,\label{lpin}
\end{equation}
and introduce a vector meson field as a gauge boson of the hidden local symmetry
\begin{equation}
i \slashed \partial \longrightarrow i \slashed\partial - g \slashed \rho,
\end{equation}
to obtain the Lagrangian for the $\rho N \rightarrow \pi N$ vertex (see Ref~\cite{us2} for more details)
\begin{eqnarray}\nonumber
\mathcal{L}_{\pi N \rho N} &=& 
- i \frac{g }{2 f_\pi} \bar{N}   \left[ \pi ,  \rho^\mu \right]  \gamma_\mu \gamma_5 N\\
&\rightarrow& - i \frac{g g_A}{2 f_\pi} \bar{N}   \left[ \pi ,  \rho^\mu \right]  \gamma_\mu \gamma_5 N,\label{lpnrn}
\end{eqnarray}
where   $\pi = \vec{\tau} \cdot \vec{\pi}$, $\rho = \vec{\tau} \cdot \dfrac{\vec{\rho}}{2}$ and $f_\pi =$ 93 MeV is the pion decay constant.

The  SU(3) Lagrangian corresponding to Eq.~(\ref{lpnrn}) can be written as 
\begin{eqnarray}
\mathcal{L}_{PBVB} = \frac{-i g}{2 f_\pi} \left ( \tilde{F} \langle \bar{B} \gamma_\mu \gamma_5 \left[ \left[ P, V_\mu \right], B \right] \rangle + 
\tilde{D} \langle \bar{B} \gamma_\mu \gamma_5 \left\{ \left[ P, V_\mu \right], B \right\}  \rangle \right), \label{pbvb}
\end{eqnarray}
where the trace $\langle ... \rangle$ has to be calculated in the flavor space and $\tilde{F} = 0.46$, $\tilde{D}=0.8$ such that  $\tilde{F} + \tilde{D} \simeq  g_A = 1.26$. The ratio 
$\tilde{D}/(\tilde{F} + \tilde{D}) \sim 0.63$ here is close to the quark model value of 0.6, and the empirical values of $\tilde{F}$ and $\tilde{D}$ can be found, for example, in Ref.~\cite{Yamanishi:2007zza}.
Using Eq.~(\ref{pbvb}), amplitudes for all the processes within SU(3) have been obtained and are given in Ref.~\cite{us2}.

In the next section, we will discuss some of the most important results of our work on this topic.

\section{Results and discussions}
\subsection{Strangeness $0$ resonances found in the VB systems}
Having calculated the amplitudes for different VB coupled channels with total strangeness 0, which are $\rho N$, $\omega N$, $\phi N$, $K^* \Sigma$ and $K^* \Lambda$, we solve  the Bethe-Salpeter equation
\begin{equation}
T = V + V G T, \label{bs}
\end{equation}
where the kernels $V (= V^t + V^{CT}  + V^u + V^s)$ are effectively treated as contact interactions, which, in turn, can be factorized out of the loop integral involved
in Eq.~(\ref{bs}). Such a treatment, which is suitable for studies involving low-energy interactions, reduces solving of integral Eq.~(\ref{bs}) to an algebraic one.

 Since we are looking for dynamical generation of resonances, which implies working with low-energy dynamics, we calculate all the amplitudes in the s-wave. This means that the states found in our work can have $J^\pi = 1/2^-, 3/2^-$. Further, the total isospin of the VB system can be 1/2 or 3/2. Let us  first discuss the results obtained for the total isospin 1/2 case. In Fig.~\ref{fig2}, we show the squared amplitude for the $\rho N \rightarrow \rho N$ process. We find it very illustrative to show the results obtained by considering the $t$-channel exchange only for the VB interaction and  compare them with those obtained by considering all the four diagrams of Fig.~\ref{fig1}.

 From  Fig.~\ref{fig2}, one can clearly see that the  $\rho N$  amplitudes obtained from the $t$-channel are identical for spin 1/2 and 3/2 (upper-left and right panel), as expected from a spin independent interaction. However, a drastic change is seen when the amplitudes are obtained by solving Eq.~(\ref{bs}) with the sum of the $s$-, $t$-, $u$-channels and the contact interaction as the kernel. In this case no peak is seen for the spin 1/2 configuration (see the lower left panel)  while a more pronounced peak is found for spin 3/2 (shown in the lower right panel). Thus the spin degeneracy of the $t$-channel
amplitude gets lifted by adding the spin dependent contact interaction and the $s$-, $u$-channel diagrams.  To identify the peak seen in the spin 3/2 amplitude with a known resonance, we go to the complex plane and find a pole at $1637 - i35$~MeV. We find it to couple strongly to the $\rho N$ and $K^* \Lambda$ channel. These properties are  very similar to those of the $N^* (1700)~D_{13}$ resonance listed by the particle data group (PDG) \cite{pdg}. Our findings are also in good agreement with other independent phenomenological studies  \cite{michael,alwin1,alwin2}.

\begin{figure}[ht]
\begin{tabular}{cc}
\includegraphics[width=0.4\linewidth, height=4cm]{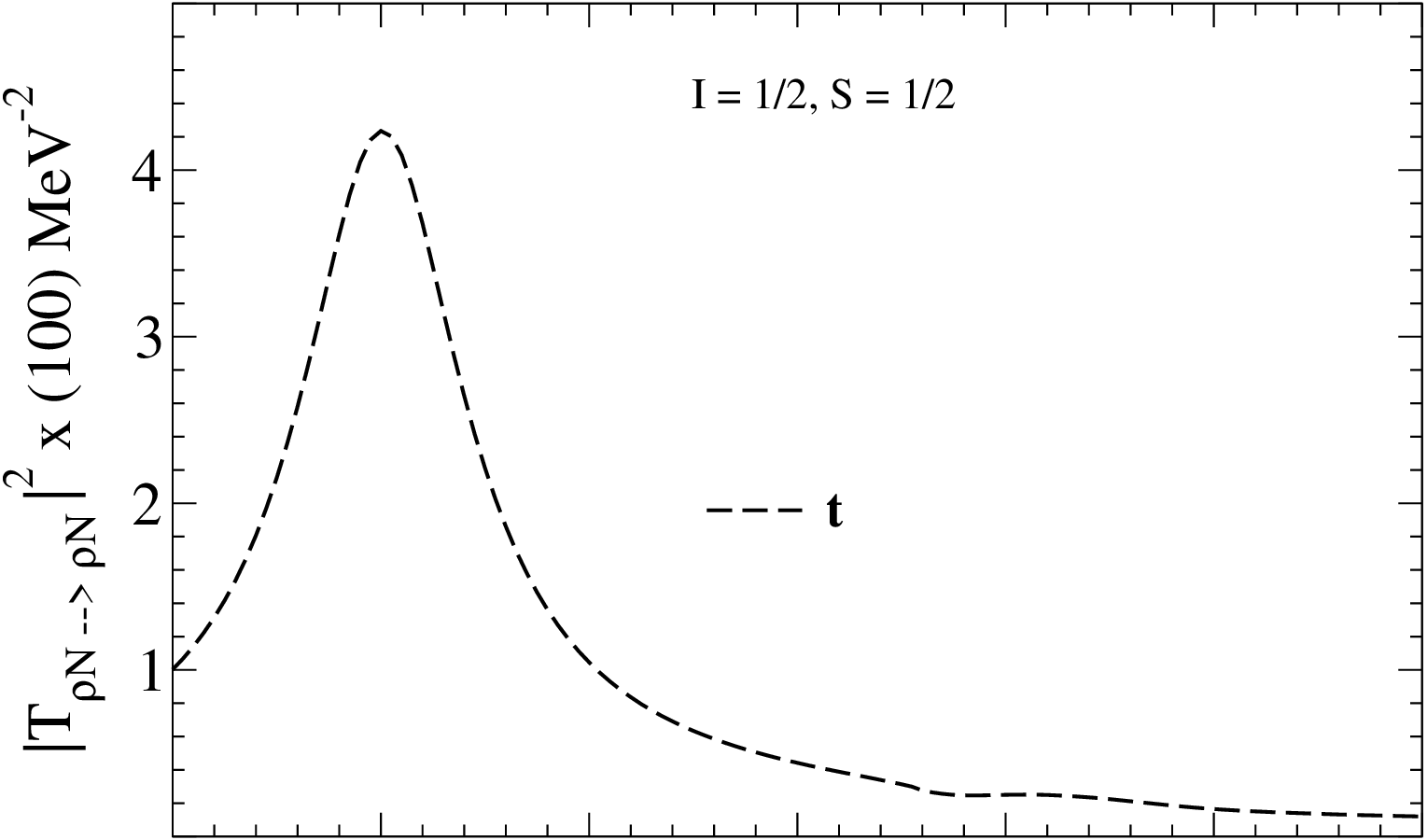}&\hspace{0.04cm}
\includegraphics[width=0.4\linewidth, height=4cm]{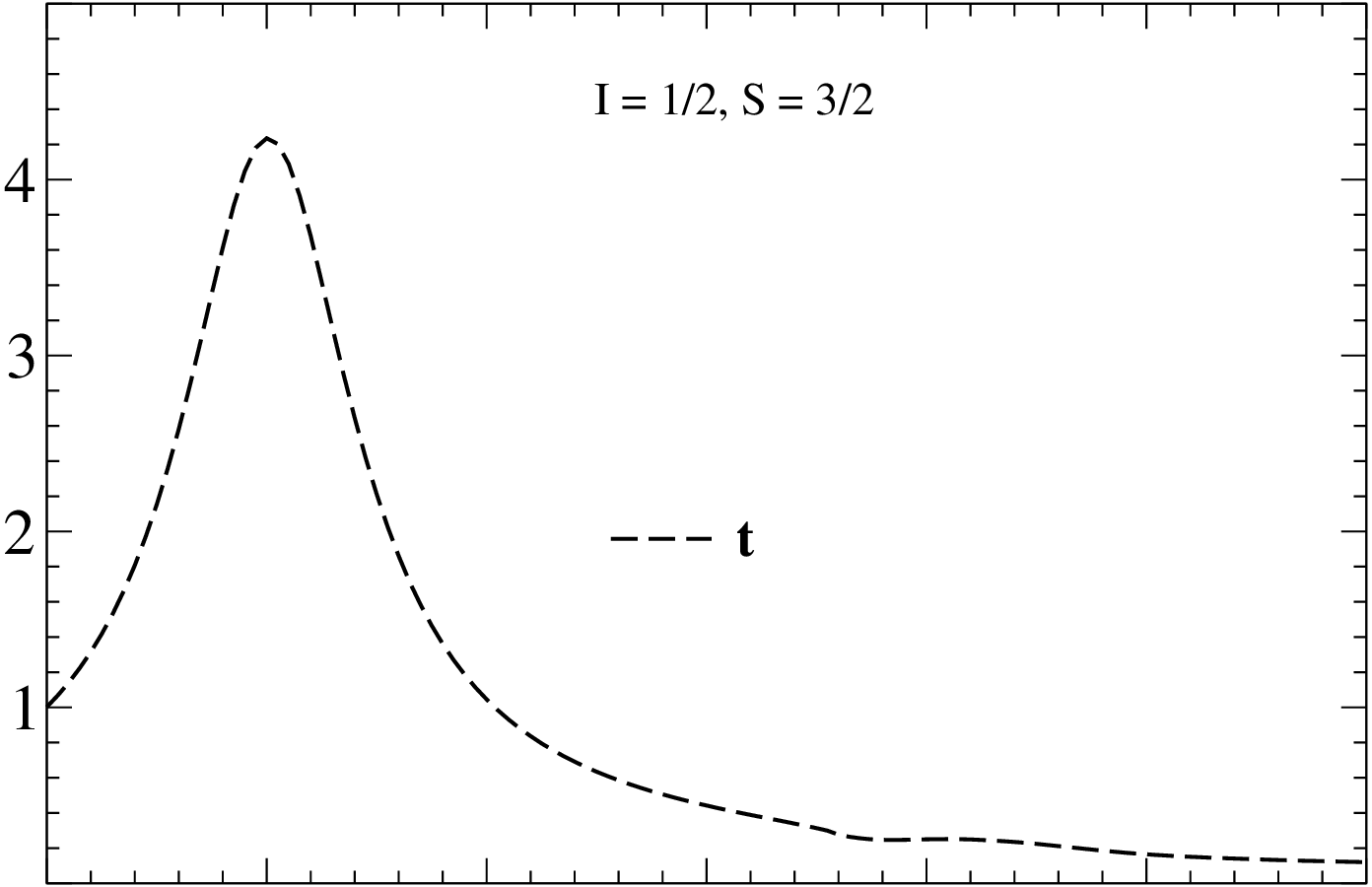}\\[-0.17cm]
\includegraphics[width=0.4\linewidth,height=4.5cm]{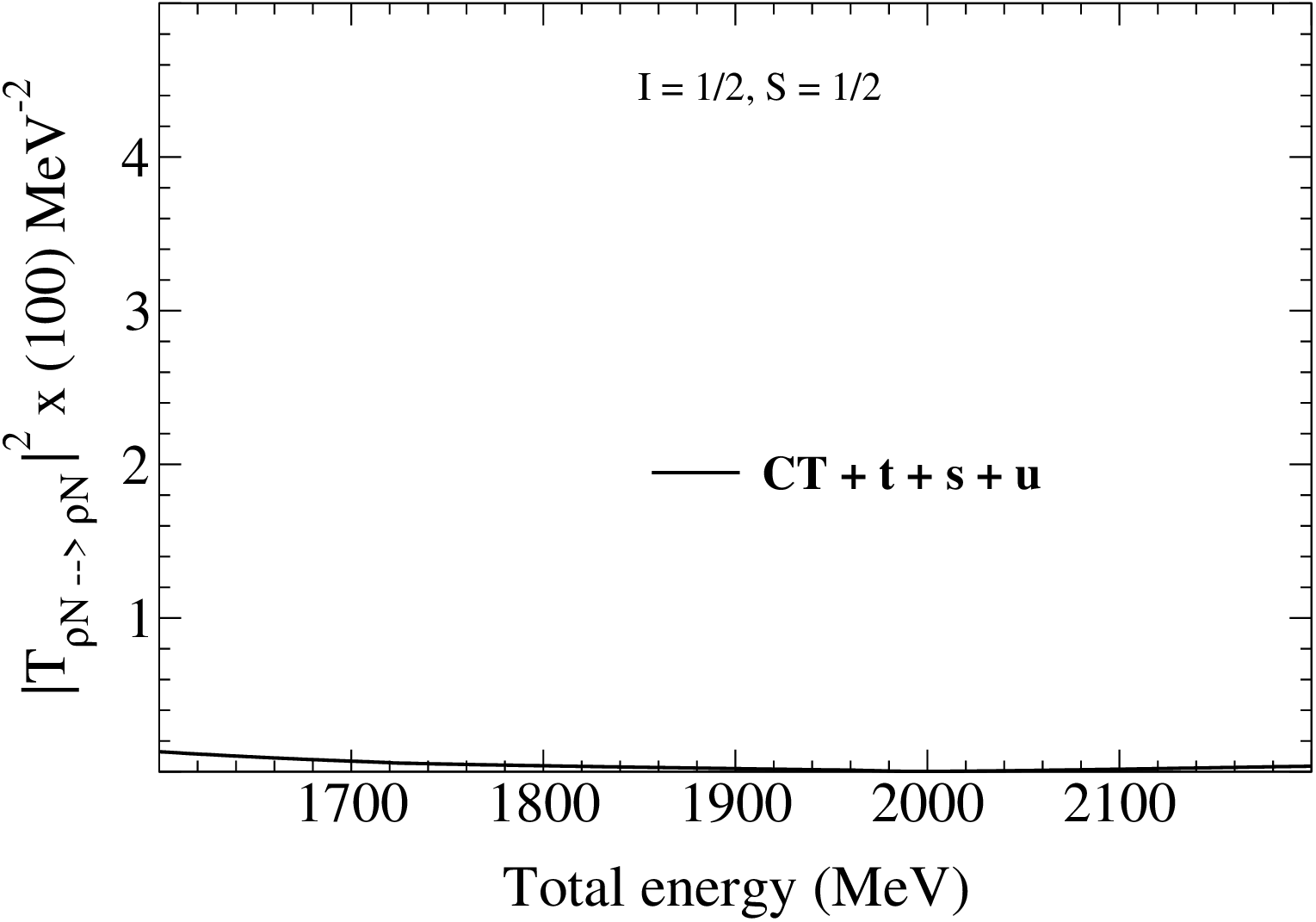}&
\includegraphics[width=0.41\linewidth,height=4.5cm]{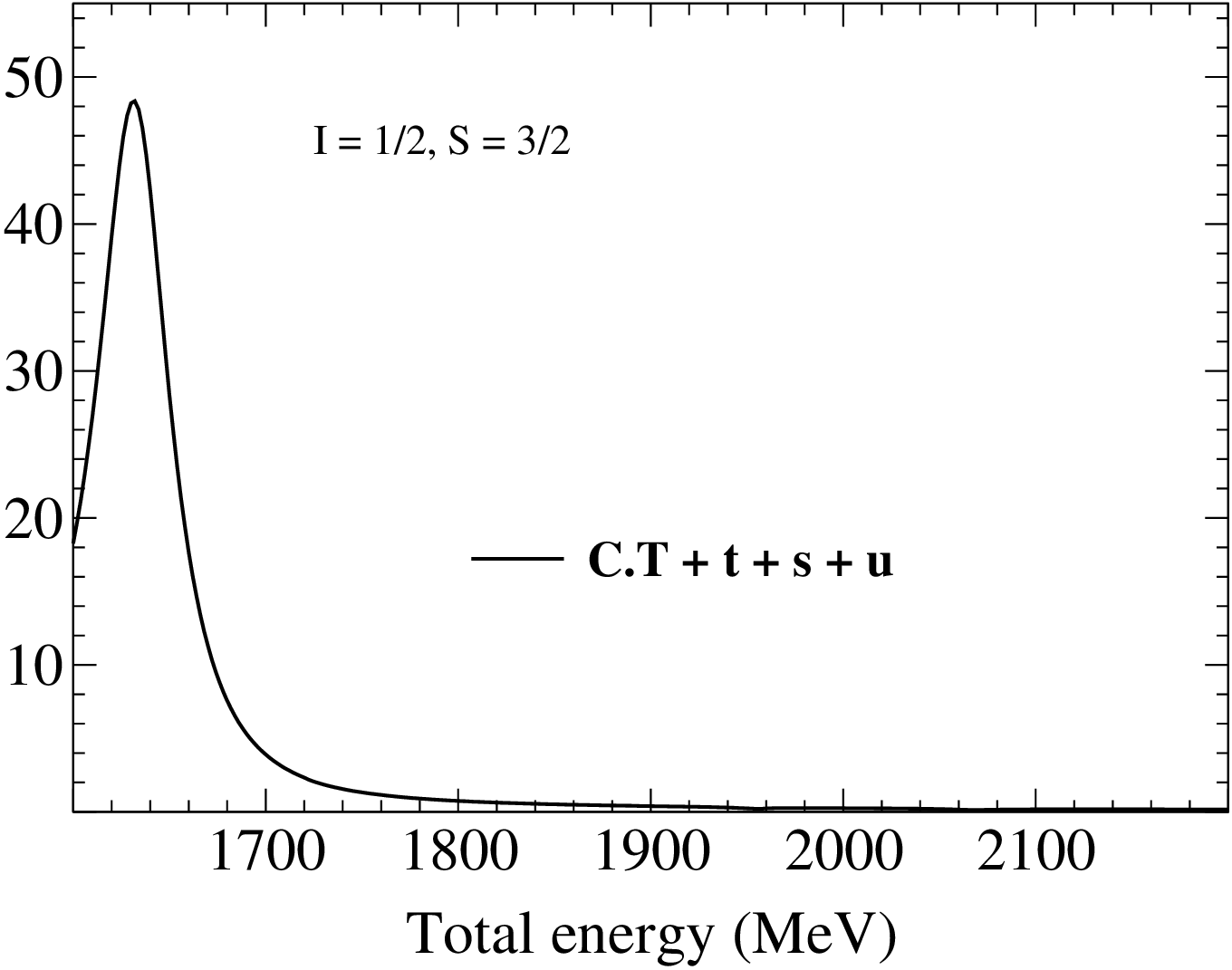}
\end{tabular} 
\caption{\label{fig2} Squared amplitudes for the $\rho N \rightarrow \rho N$ process in the isospin 1/2 configuration. The upper panels show the results obtained by
considering the $t$-channel diagram for VB interactions whereas the lower panels show the effect of taking all the four diagrams of Fig.~\ref{fig1} into account.  }
\end{figure}

We find similar changes in all the coupled channels. In Fig.~\ref{fig3}, we show the square amplitude for another channel: $K^* \Sigma$. In this case, once again,
the upper-left  and -right panels of Fig.~\ref{fig3} depict the same amplitudes for both spin 1/2 and 3/2, which show a peak near 2 GeV. We found that these peaks corresponded to a double pole structure in the complex plane \cite{us1}. The lower panels show the results obtained by considering all four
diagrams of Fig.~\ref{fig1}. In this case, both the states in spin 1/2 and 3/2  survive, although the peak positions differ by about 100 MeV.  We find that the peak in the spin 3/2 amplitude corresponds to a pole in the complex plane at $2071- i70$~MeV. This pole is found to couple strongly to the $\omega N$, $\phi N$ and $K^* \Lambda$ channels and can be related to the $N^* (2080)~D_{13}$. In the spin 1/2 case, we find a pole at $1977 - i22$~MeV. There is no known $J^\pi = 1/2^-$ state in Ref.~\cite{pdg} with a mass very close to this pole, however, an enhancement of the cross section is seen near 2 GeV
region in the photo-production of the $\phi$-meson on a nucleon studied by the LEPS group \cite{leps}. This enhancement could possibly be explained in terms of the 
spin $1/2^-$ resonance found
in our work. In fact, a phenomenological analysis of the $\gamma N \rightarrow \phi N$ reaction showed that a
better fit to the LEPS data \cite{leps} was found when a $J^\pi = 1/2^-$ resonance was included in
their study as compared to the one obtained by including a $1/2^+$ resonance \cite{ozakisan}. 

We have not discussed the results for total isospin 3/2 so far, where we also find very important contribution from all the diagrams shown in Fig.~\ref{fig1}. The $t$-channel interaction in isospin 3/2 is repulsive (for spin 1/2 as well as 3/2) and, thus, does not give rise to formation of any states.
\begin{figure}[ht]
\begin{tabular}{cc}
\includegraphics[width=0.4\linewidth, height=4cm]{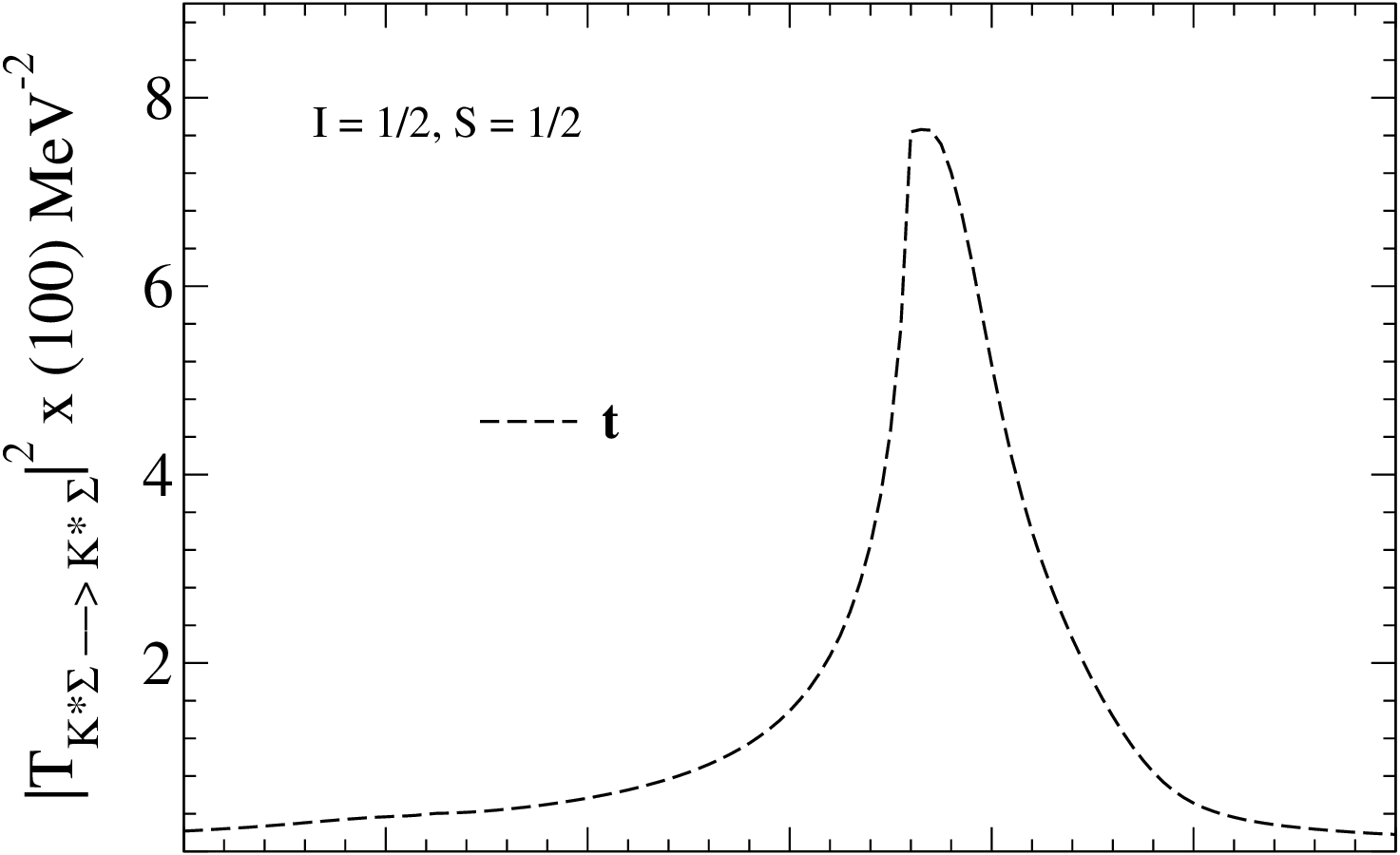}&\hspace{0.04cm}
\includegraphics[width=0.4\linewidth, height=4cm]{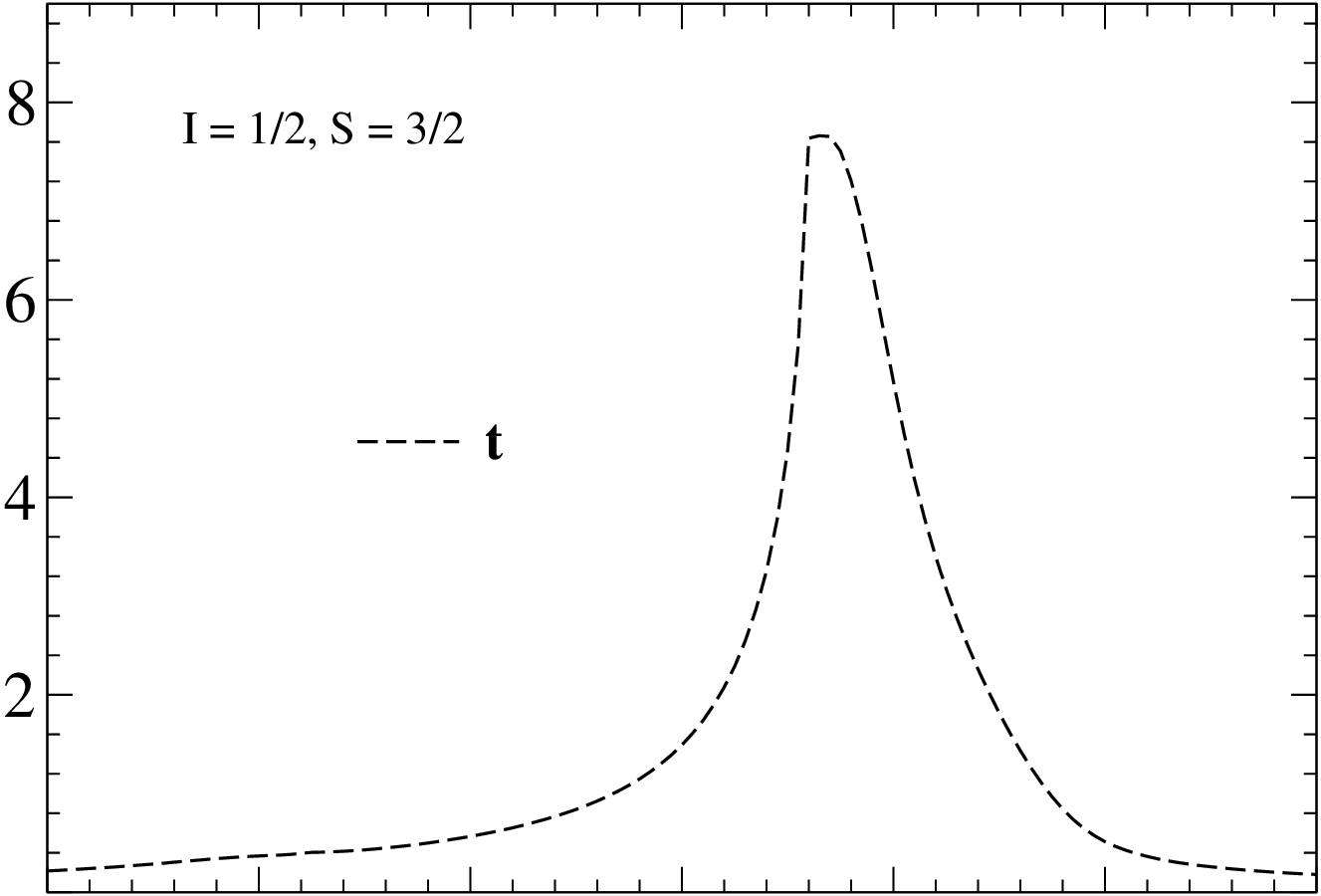}\\[-0.18cm]
\includegraphics[width=0.4\linewidth,height=4.5cm]{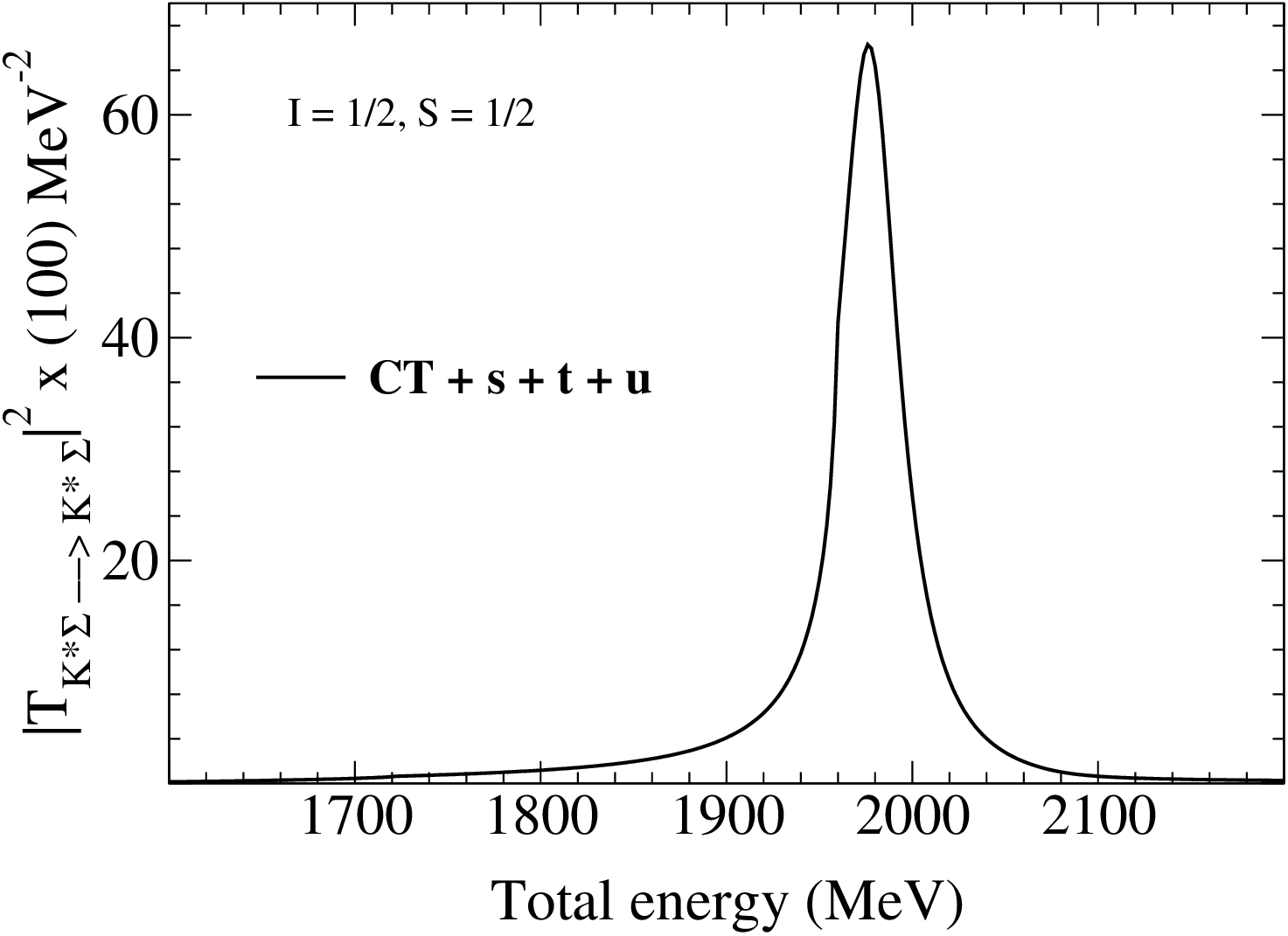}&
\includegraphics[width=0.41\linewidth,height=4.5cm]{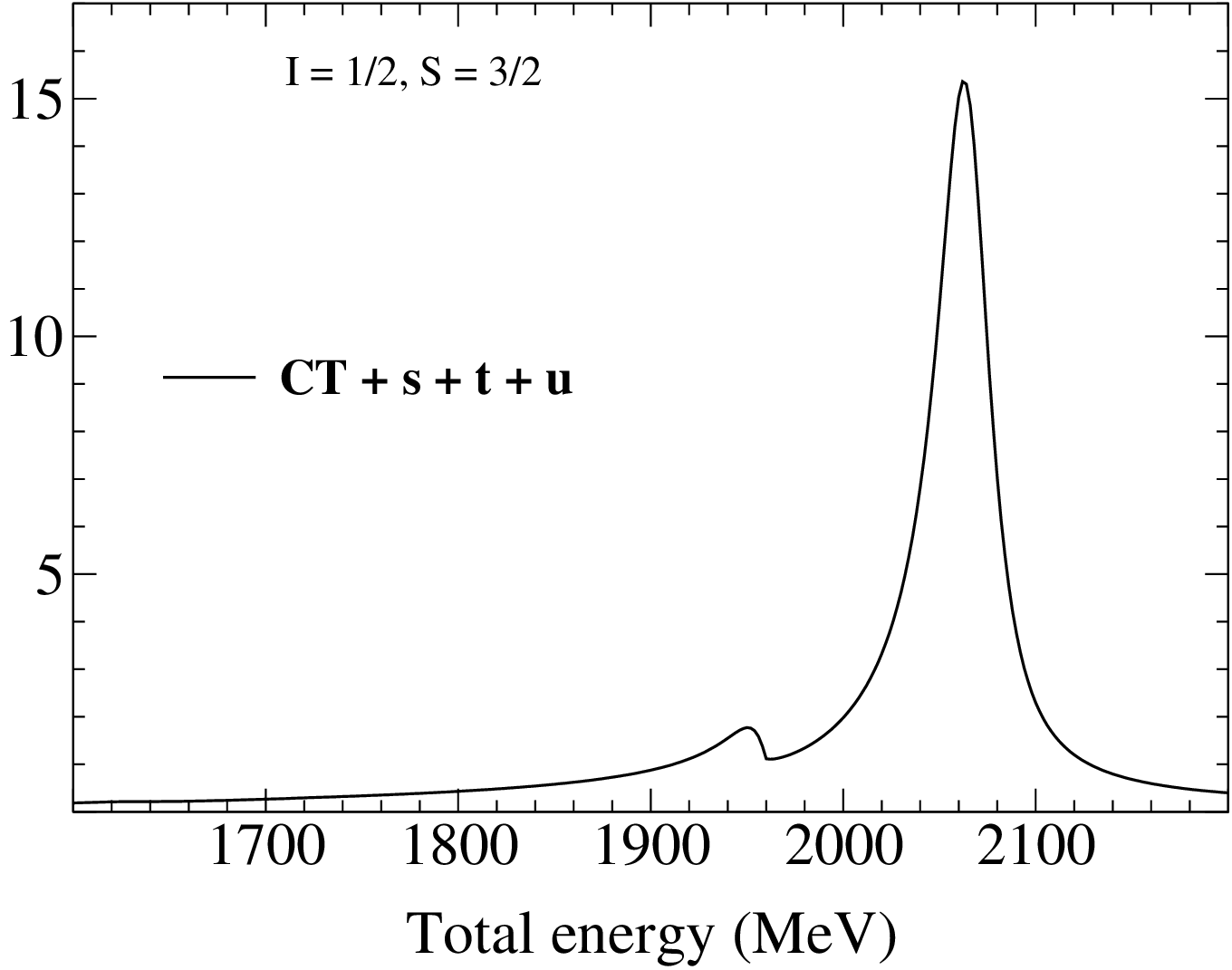}
\end{tabular} 
\caption{\label{fig3} Same as Fig.~\ref{fig2} but for the  $K^* \Sigma \rightarrow K^* \Sigma$ process.}
\end{figure}
 However, the vector meson-baryon contact interaction is strongly attractive for isospin 3/2 and spin 1/2, which leads to the formation of a resonance near 2 GeV.  In Fig.~\ref{fig4}  we show the squared amplitude for the isospin 3/2 VB channels, to which only $\rho N$ and $K^* \Sigma$ contribute. The left panel 
of this figure shows the amplitudes obtained by considering  the $t$-channel interaction only while the right panel shows the results obtained by summing the $s$-, $t$-, $u$- channel exchanges, and the  contact interaction. A pole was found at $2006 -i112$~MeV corresponding to the peak  shown in the right panel.  This state can be related with 
$\Delta (1900)~S_{31}$.
\begin{figure}[ht]
\begin{tabular}{cc}
\includegraphics[width=0.4\linewidth]{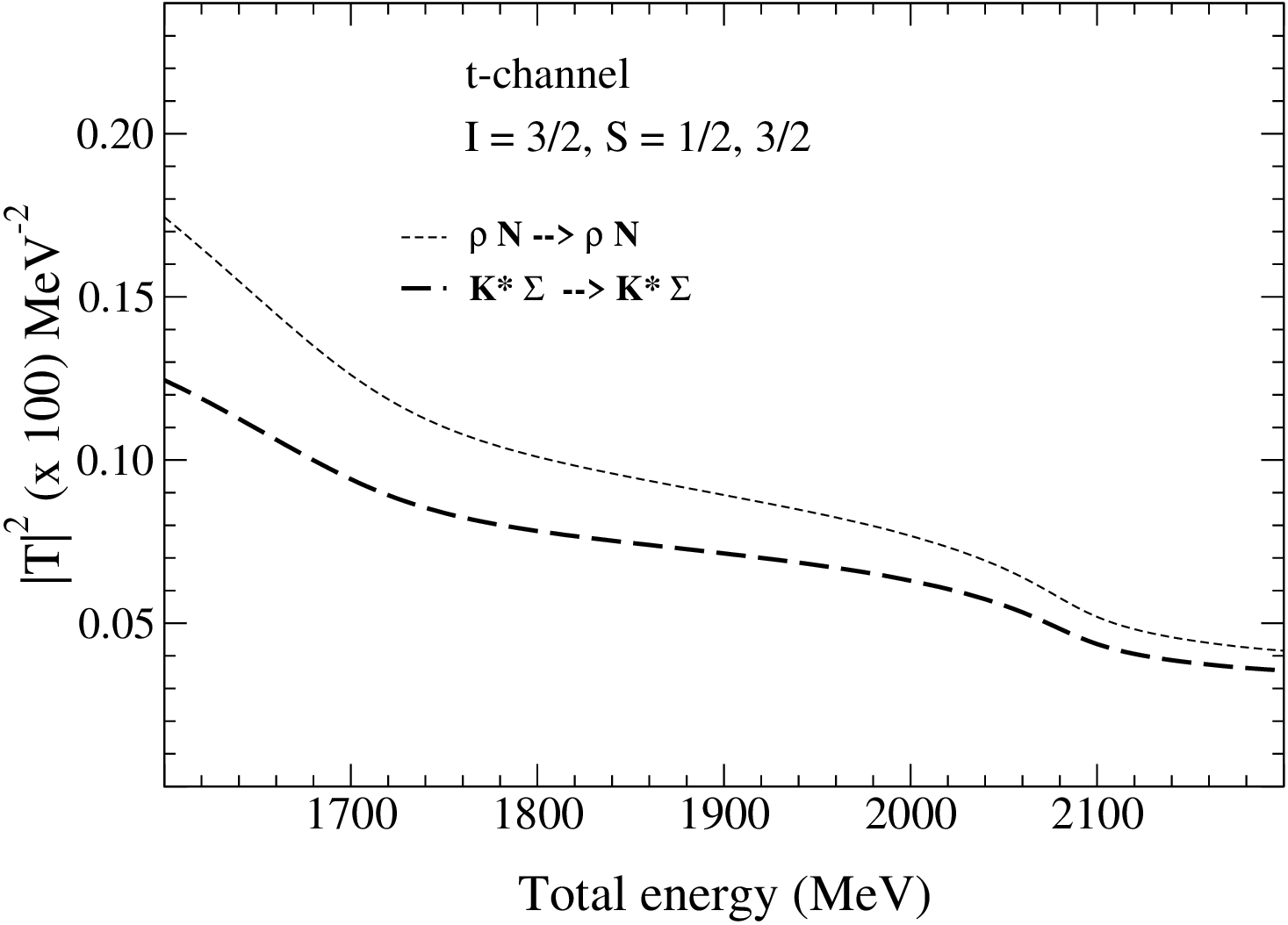}&\hspace{0.04cm}
\includegraphics[width=0.4\linewidth]{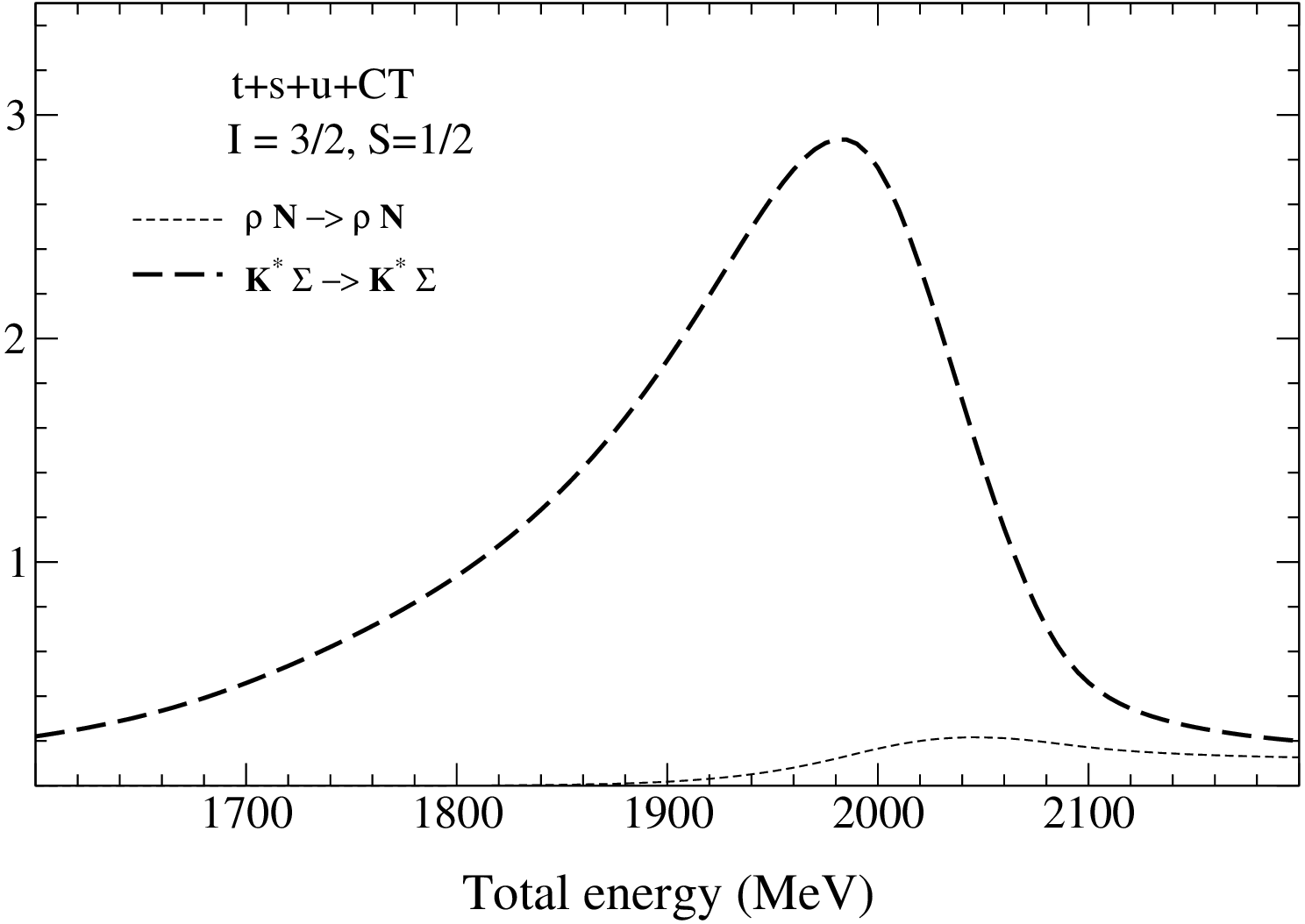}
\end{tabular} 
\caption{\label{fig4} Squared amplitude for VB channels with total strangeness zero in the isospin 3/2 configuration.}
\end{figure}

\subsection{Strangeness $-1$  PB-VB coupled systems}

Strangeness $-1$ is probably the most explored sector when it comes to discussing  the dynamically generated resonances. Starting from the work of Dalitz \cite{dalitz} to more recent studies reported in Refs.~\cite{dgr2,dgr3,dgr4,jido2pole,hyodo1,hyodo2}, strangeness $-1$ systems have been the center of attention when looking for resonances. There are at least two low-lying resonances which are good candidates for qualifying as the dynamically generated resonances in PB systems. However, these resonances have not been coupled to the vector meson-baryon systems. We made such a study within the formalism explained in the previous section (see  Ref.~\cite{us2} for more details). We found that 
the known properties of these resonances, i.e., the corresponding pole positions, their coupling to different PB channels and the PB amplitudes on the real axis, did not change much by coupling the VB channels. However, we found that the low-lying resonances couple very strongly to the heavy, closed VB channels. It is important to mention here that the large couplings of the low-lying $\Lambda$'s to the VB channels found in our work do not imply the presence of a large fraction of the VB component in the wave function of these resonances since the large mass difference between the two would suppress it. Therefore, the interpretation of the low-lying $\Lambda$'s as PB molecular states does not change. However, our findings could
have some implications on, for example, the photoproduction of the $\Lambda$ resonances where the production mechanism proceeding through the exchange of a  vector meson could become important \cite{Nam:2008jy}. This should be verified in future.

We show the pole positions for the low-lying resonances and their couplings to the different PB and VB channels for both uncoupled as well as coupled case (which corresponds to the KR coupling $6$ between the PB and VB channels \cite{us2}) in Table~\ref{iso0_1}. As can be seen from this table, the couplings of the poles to the VB channels is null when the latter ones are not coupled to PB, while a very large coupling gets developed on coupling both systems. The larger couplings of the poles to  various VB channels 
have been highlighted by boldfacing them in Table~\ref{iso0_1}.

\begin{table}[htbp]
\caption[]{ $g^{i}$ couplings of low-lying resonances to the PB and VB channels for different strengths of  the coupling between PB-VB systems: $0$ indicates no coupling between the two and $6$ indicates fully coupled PB-VB systems (see \cite{us2} for more details).}  \label{iso0_1}
\begin{tabular}{l|cccc|cc}
\hline
& \multicolumn{4}{c}{$\Lambda(1405)$} &\multicolumn{2}{c}{$\Lambda(1670)$}\\
& \multicolumn{2}{c}{Pole1} &\multicolumn{2}{c}{Pole2}&\multicolumn{2}{ c}{}\\
\hline
PB-VB&$0$ &$ 6$&$ 0$&$6$&$ 0$&$ 6$\\
coupling &&&&&&\\
\hline
$M_R - i\Gamma/2 \rightarrow$ &$1377 - i63$ &$1357 - i53$&$1430 - i15$&$1412 - i11$&$1767 - i25$&$1744 - i28$\\
(MeV)&&&&&&\\
\hline
Channels $\downarrow$& \multicolumn{6}{c}{Couplings ($g^i$) of the poles to the different channels} \\
\hline\hline
$\bar{K} N$             &$~1.4 - i 1.6$ &$~1.1 - i 1.4$&$~2.4 + i 1.1$ &$~2.8 + i 0.5$& $~0.2 - i 0.5$&$~0.3 - i 0.6$\\
$\pi \Sigma$           &$-2.3 + i 1.4$ &$-2.2 + i 1.4$&$-0.2 - i 1.4$ &$-0.2 - i 1.1$& $~0.1 + i 0.2$&$~0.1 + i 0.3$\\
$\eta \Lambda$      &$~0.2 - i 0.7$ &$~0.1 - i 0.6$&$~1.3 + i 0.3$&$~1.5 + i 0.1$&$-1.0 + i 0.3$ &$-1.0 + i 0.3$\\
$K \Xi$                      &$-0.4 + i 0.4$ & $-0.6 + i 0.4$&$~0.0 - i 0.3$&$~0.0 - i 0.3$&$~3.2 + i 0.3$ &$~3.4 + i 0.2$\\
$\bar{K^*}N$            &$~0.0 + i 0.0$ &$-1.7 + i 0.7$&$~0.0 + i 0.0$ &${\bf-0.1 - i 5.3}$&$~0.0 + i 0.0$ &$-0.3 + i 1.1$\\
$\omega \Lambda$&$~0.0 + i 0.0$ &$-0.7 - i 0.3$&$~0.0 + i 0.0$ &$~0.2 - i 1.8$&$~0.0 + i 0.0$ &$~0.1 - i 0.1$\\
$\rho \Sigma$          & $~ 0.0 + i 0.0$&${\bf~1.3 + i 6.8}$&$~0.0 + i 0.0$ &${\bf-2.4 - i 1.6}$&$~0.0 + i 0.0$ &${\bf~0.3 - i 3.5}$\\
$\phi \Lambda$       &$~0.0 + i 0.0$ &$~1.0 + i 0.5$&$~0.0 + i 0.0$ &$-0.3 + i 2.6$&$~0.0 + i 0.0$ &$-0.2 + i 0.1$\\
$K^*\Xi$                    & $~0.0 + i 0.0$&${\bf~1.3 + i 5.7}$&$~0.0 + i 0.0$ &$-2.0 - i 0.5$&$~0.0 + i 0.0$ &$~0.5 - i 1.2$\\
\hline
\end{tabular}
\end{table}

Further, although our purpose was to study low-lying resonances first, for which we relied on  the $t$-channel exchange to calculate PB $\rightarrow$ PB and VB $\rightarrow$ VB amplitudes, to 
draw any concrete conclusion about higher mass resonances, we should take into account a more complete VB $\rightarrow$ VB interactions as shown in Ref.~\cite{us1}. Still, within the present formalism, we could test  if the widths of the resonances in the 1800-2100 MeV region increased a lot by coupling the lower mass (PB) open channels. We found that the higher mass resonance poles did not always attain larger widths when coupling them to PB channels, sometimes poles disappeared and other times new poles appeared by coupling VB to the light PB channels \cite{us2}. For instance, we found that a new pole gets generated near 1440 MeV in total isospin 1 case when PB and VB are coupled. This pole can be related to $\Sigma(1480)$ \cite{pdg,zychor,zou1400}.  However, to make stronger claims on the properties of the resonances found  in 1800-2100 MeV, we must carry out a more detailed calculations by obtaining the VB amplitudes from the different diagrams shown to be important in Ref.~\cite{us1}. Such a study is being carried out at this moment and we expect it to finalize soon.

\section{Summary and outlook}
We can summarize our study on the baryon resonances coupled to vector meson-baryon systems as follows:
\itemize
\item{A gauge  invariant Lagrangian under the hidden local symmetry gives rise to significantly important tree-level contributions from the $s$-, $t$- and $u$- channel exchange diagrams and a contact term. Our findings show that none of these contributions should be assumed to be negligible {\it a priori}.} 
\item{The resulting vector meson-baryon interaction is very spin-isospin dependent. This is something which should be expected when two particles with similar mass and non-zero spin interact.}
\item{ Many low-lying resonances like $\Lambda$(1405) couple strongly to VB channels, which is a very useful information, for instance, to study photoproduction of $\Lambda$(1405).}
\item{When heavier VB resonance poles are allowed to couple to PB channels,  they do not always get wider. Sometimes they can  disappear. 
Also new poles can appear sometimes. For example, a new isospin 1 resonance is found  to arise due to the PB-VB coupled dynamics near 1430 MeV}

Finally, as an outlook we would like to say that a PB-VB coupled channel calculation should be made with a more detailed VB interaction.
Further, it  is important to use these amplitudes to study relevant reactions, which have been studied experimentally.

\section{Acknowledgments}
This work is partly  supported  by the Grant-in-Aid for Scientific Research on Priority Areas titled ÒElucidation of New Hadrons with
a Variety of Flavors" (E01: 21105006 for K.P.K and A.H) and (22105510 for H.~N) and the authors acknowledge the same. A.~M.~T  
is thankful to the support from the Grant-in-Aid for the Global COE Program ÒThe Next
Generation of Physics, Spun from Universality and EmergenceÓ from the Ministry of Education,
Culture, Sports, Science and Technology (MEXT) of Japan.

\end{document}